\documentclass[aip,reprint,jcp,amsmath,amssymb,superscriptaddress]{revtex4-1}
\usepackage{graphicx}
\usepackage{color}
\usepackage[normalem]{ulem}
\newif\ifHighlitedChanges
\def\ifHighlitedChanges{\iftrue}
\ifHighlitedChanges
  
  \def\STRIKE#1{{\color{red}\sout{#1}}}
\else
  
  \def\STRIKE#1{\relax}
\fi
\newcommand{\lu}{\lambda_\text{u}}
\newcommand{\ls}{\lambda_\text{s}}
\newcommand{\omb}{\omega_\text{b}}

\newcommand{\kT}{k_{\text B}T}
\newcommand{\vmin}{v_\text{min}}

\let\Re\relax
\DeclareMathOperator{\Re}{Re}

\def\env@matrix{\hskip -\arraycolsep
  \let\@ifnextchar\new@ifnextchar
  \array{*\c@MaxMatrixCols c}}	
	
\makeatletter
\renewcommand*\env@matrix[1][c]{\hskip -\arraycolsep
  \let\@ifnextchar\new@ifnextchar
  \array{*\c@MaxMatrixCols #1}}
\makeatother
\begin{document}
\bibliographystyle{apsrev}
\title{Chemical reactions induced by oscillating external fields in weak thermal environments}
\author{Galen T. Craven}
\affiliation{Center for Computational Molecular Science and Technology, \\
School of Chemistry and Biochemistry, \\
Georgia Institute of Technology, \\
Atlanta, GA  30332-0400}

\author{Thomas Bartsch}
\affiliation{Department of Mathematical Sciences, \\
Loughborough University, \\
Loughborough LE11 3TU, \\
United Kingdom}

\author{Rigoberto Hernandez}
\thanks{Author to whom correspondence should be addressed}
\email{hernandez@chemistry.gatech.edu.}
\affiliation{Center for Computational Molecular Science and Technology, \\
School of Chemistry and Biochemistry, \\
Georgia Institute of Technology, \\
Atlanta, GA  30332-0400}

\begin{abstract} 
Chemical reaction rates must increasingly be determined in systems 
that evolve under the control of external stimuli.
In these systems,
when a reactant population is induced to cross an energy barrier 
through forcing from a temporally varying external field,
the transition state that the reaction must pass through 
during the transformation from reactant to product is no longer a fixed geometric structure, 
but is instead time-dependent. 
For a periodically forced model reaction, we develop a recrossing-free dividing surface that is attached to a transition state trajectory 
[T. Bartsch, R. Hernandez, and T. Uzer, 
Phys. Rev. Lett. {\bf 95}, 058301 (2005)].
We have previously shown that for single-mode sinusoidal driving, 
the stability of the time-varying transition state directly determines the reaction rate
 [G. T. Craven, T. Bartsch, and R. Hernandez, 
J. Chem. Phys. \textbf{141}, 041106 (2014)].
Here, we extend our previous work to the case of multi-mode driving waveforms.
Excellent agreement is observed between the rates predicted by 
stability analysis and rates
obtained through numerical calculation of the reactive flux.
We also show that the optimal dividing surface 
and the resulting reaction rate
for a reactive system driven by weak thermal noise
can be approximated well using the transition state geometry of the
underlying deterministic system.
This agreement persists as long as
the thermal driving strength 
is less than the order of that of the periodic driving.
The power of this result is its simplicity.
The surprising accuracy of the time-dependent
noise-free geometry for obtaining
transition state theory rates
in chemical reactions driven by periodic fields
reveals the dynamics without requiring the cost of
brute-force calculations.
\end{abstract}

   \maketitle
\section{\label{sec:Intro}Introduction}

Optimal control of reaction pathways in systems undergoing 
configurational changes can be achieved through forcing from tailored external fields. 
These fields can be tuned to induce specific deformations on a potential energy surface, providing
control of state-to-state transitions. \cite{Yamanouchi2002,Sussman2006,Kawai11laser}
In these processes, a formally exact classical rate calculation 
can be obtained through modern-day transition state theory (TST).
\cite{truh83,truh84,mill93,Truhlar96,Waalkens2008,hern10a}
The principal assumptions of TST are that 
(1) the distribution of energy states in 
the reactant configuration, and at the TS, are given by equilibrium distributions
and (2) there exists a hypersurface 
between reactant and product confirmations 
that is crossed only once by reactive trajectories during the  traversal of a free energy barrier separating these basins.
The TST reaction rate is calculated from the flux through this dividing surface (DS).  
If the DS is recrossed by reactive trajectories, 
TST will give an overestimate to the classical reaction rate.
Determination of a recrossing-free DS
therefore leads to classically exact TST rates systems 
in which equilibrium statistical mechanics is applicable.

A phase space DS that is free of recrossings can be constructed in conservative systems at energies 
close to the reaction threshold.
In systems with two degrees of freedom, the optimal DS is the configuration space projection of an
unstable periodic orbit (PO).\cite{pollak78,pollak79,pech79a,pollak80} 
In systems with higher dimensionality, the generalization of this PO
is a normally hyperbolic invariant manifold (NHIM).\cite{hern93b,hern94,Uzer02,deLeon2,Li06prl,Waalkens04b,Ezra2009,Ezra2009a,Teramoto11} 
The NHIM bounds the TS, being one less in dimension.\cite{Uzer02}
It defines a recrossing-free surface at energies below bifurcation thresholds. \cite{Inarrea2011,Allahem12,Mackay2014}
Reactive trajectories are mediated by stable and unstable manifolds (reaction pathways) attached to the NHIM.
These pathways persist even in reactions 
whose state-to-state transitions 
are not dictated by purely configurational changes.\cite{Waalkens13}

In systems subjected to time-varying external forcing,
the characterization of the NHIM as a hypersphere of constant energy breaks down.
For example, field-matter interactions 
constitute processes in which energy is exchanged with 
a reacting system.  
These interactions lead to emergent and controllable behavior in 
assembly phenomena,\cite{Elsner09,Jager11,Prokop12,Ma13} 
organic synthesis,\cite{Lids01} 
protein folding,\cite{Gruebele2014} 
the detection of DNA,\cite{Loget11}
and photodissociation.\cite{Corrales2014}
Knowledge of the mechanism by which these interactions mediate reactive flow provides a methodological tool
in the design of molecular devices with unique functionality.\cite{Saha2007,Michl2009,Zazza2013}

Materials that 
undergo conformational changes in response to an external trigger 
offer examples of such emergent technology. \cite{Browne2006,Kay2007,Balzani2007,Michl2009,Meng2013}
Stimuli such as thermal variations, electric fields, and photoinduction have been used as
triggers for the conversion of chemical energy into mechanical work. \cite{Fer2000,Leigh2003,Fletcher2005,Zazza2013}
Assemblies that convert chemical energy into directional motion 
can achieved through isomerization reactions which are induced either from light or applied electric fields. \cite{Horinek2005,Klok2008,Saha2007} 
In these responsive materials, controlling the rate and pathway at which reactants transform to products 
is fundamental to harness mechanical actions for applicative purposes.

The aim of this paper is to develop a rate theory for reactions that are driven by 
periodic external fields in weak thermal environments.
In the absence of noise, a dissipative system that is 
periodically driven admits a DS that is free of recrossings.\cite{hern14b}
This structure differs from the canonical view of the TS 
wherein the TS is a structure fixed in time at a saddle point 
on the potential energy surface.
Here, we develop a rate theory based on reactive flux through this 
recrossing-free DS and the stability of the corresponding TS. 
In Ref.~\citenum{hern14f}, we found that the 
stability of the moving TS directly determines the reaction rate
for single-mode sinusoidal driving.
In conservative systems, stability analysis is known to characterize
molecular motions as well as determine the rate of configurational transitions. \cite{kadanoff84,skodje90,gaspard98,Green2011,Allahem12} 
Building on our previous work, 
we test the viability of stability analysis to determine reaction rates 
in systems driven by multi-mode waveforms
with no thermal driving.
The extent to which the accuracy of the rate theory
relying on the noise-free geometry 
persists in systems
that are coupled to a thermal bath is also verified through 
inclusion and variation of the thermal driving strength.

This outline of this paper is as follows: 
In Sec.~\ref{sec:Model}, a dynamical system 
is introduced to model barrier crossings 
in chemical reactions forced by periodic external fields. 
In Sec.~\ref{sec:TS_traj}, a dividing surface that is recrossing-free 
is constructed for this model in the absence of thermal driving. 
Section~\ref{sec:Rate_Theory} contains analytical theories 
to predict the reaction rates 
of driven reactions by calculation of the reactive flux through this 
dividing surface for both globally non-linear
and locally linear dynamics.
Comparison to the computational rates, computed
from numerical integration of large ensembles of trajectories,
is presented in Sec.~\ref{sec:Rates}. 
%%% HERE
Although not considered earlier,
the effect of noise on the rate
of these driven systems have also been addressed in 
Sec.~\ref{sec:Noise}.
We find that the rates computed from the noise-free geometry 
are accurate up to relatively
large values of the friction and sometimes even in the 
thermal regime.

\section{\label{sec:Model}Model Details}

%%%%%%%%%%%%%%%%%%%%%
\begin{figure*}
\includegraphics[width=17.0cm,clip]{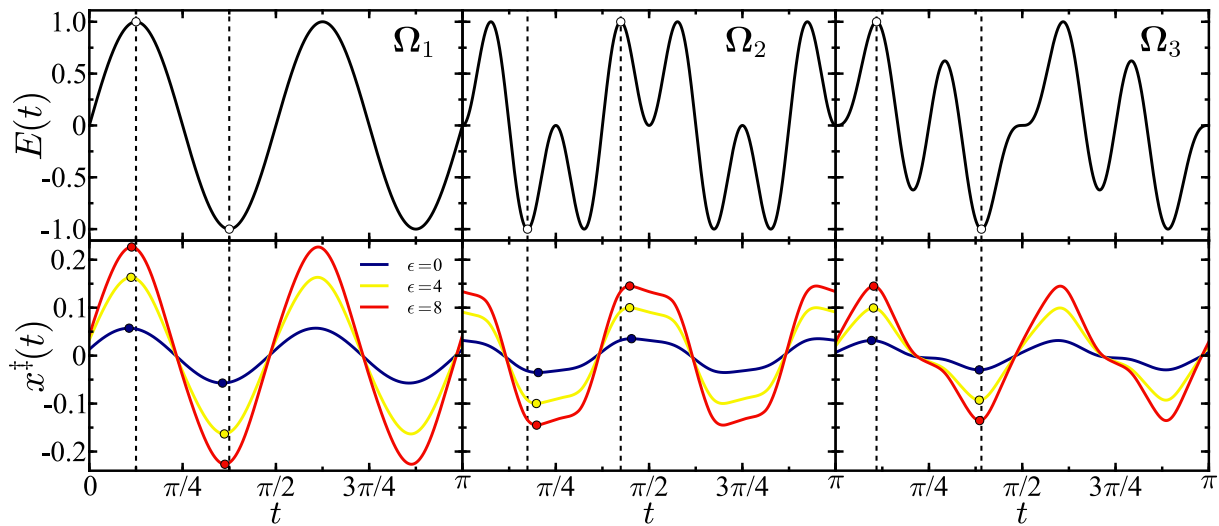}
\caption{\label{fig:TS_traj}
Functional forms of periodic driving $E(t)$ for the single $\mathbf{\Omega}_{1}$ (left),
 $\mathbf{\Omega}_{2}$ (middle), and  $\mathbf{\Omega}_{3}$ (right) 
are shown in the top row. 
The corresponding TS trajectory for each frequency set 
and anharmonic parameter values $\epsilon \in \left\{0,4,8\right\}$ are shown in the bottom row. 
Various extrema of $E(t)$ are denoted by circles with arguments shown as dashed vertical lines. 
The corresponding extrema of $x^\ddag(t)$ are denoted by circles and colored according to the respective $\epsilon$ value.
Units in all panels dimensionless.}
\end{figure*}
%%%%%%%%%%%%%%%%%%%%%

The interaction of an external field with a reactant species 
can strongly influence the mechanism 
and rate  of a reaction.\cite{Miller80,Kawai07,Kawai11laser}
As a paradigmatic example of a chemical reaction driven under kinetic control, 
we consider a particle of unit mass moving along a reaction coordinate $x$.
The trajectory of the particle begins at a position $x_0$ on the reactant side of an energy barrier 
that is moving in space under the influence of a time-dependent external field $E(t)$.
The chosen potential surface is the quartic form
\begin{equation}
	\label{eq:potanharm}
	U(x) = -\tfrac 12 \omb^2 (x-E(t))^2-\tfrac{1}{4}\epsilon(x-E(t))^4.
\end{equation}
The time dependent, instantaneous position of the moving barrier top (BT) is specified by $E(t)$.

With the inclusion of additional non-conservative dissipation 
as well as stochastic driving forces, 
a particle at a phase space point $\mathbf{\Gamma}=(x,v)$ 
moving according to the potential $(\ref{eq:potanharm})$ 
can be described by the Langevin equations of motion
\begin{equation}
	\begin{aligned}
		\label{eq:motionAnharm}
		\dot x &= v,  \\
		\dot v &= -\gamma v + \omb^2 (x -  E(t))+\epsilon (x -  E(t))^3+\sqrt{2 \sigma}\,\xi_\alpha(t),
	\end{aligned}
\end{equation}
where $\gamma \geq 0$ is a dissipation parameter, $\omb$ is the barrier frequency, 
and $\epsilon$ is an anharmonic coefficient.
Thus, for $\epsilon \neq 0$ the coordinate of the particle is non-linearly coupled to the moving barrier.
By restricting the anharmonic coefficient to values $\epsilon\geq 0$,  
there is a single maximum in the potential located at the BT.  
The random fluctuating force $\xi_\alpha(t)$ is Gaussian white noise
obeying the statistical properties
\begin{equation}
	\label{eq:noise}
	\left\langle \xi_\alpha(t) \right\rangle = 0 \;\;\; \text{and} \;\;\; \left\langle \xi_\alpha(t) \xi_\alpha(t') \right\rangle = \delta(t-t'),
\end{equation}
where $\alpha$ denotes a specific noise sequence.
The strength of the noise is varied through the parameter $\sigma \geq 0$.

Depending on the geometry of $U(x)$, initial conditions, 
as well as the specific realization of the thermal environment and the external field, 
a trajectory will either surmount the energy barrier and form product 
or remain on the reactant side. 
By calculation of the normalized flux of {\it reactive} trajectories through the 
the TS, the classical reaction rate for a system evolving through (\ref{eq:motionAnharm}) 
can be obtained.\cite{mill93}

In this article, we consider periodic external driving of the form
\begin{equation}
	\label{eq:BT}
	E(t)  = a\!\!\!\prod_{\,\,\omega\in{\mathbf{\Omega}_{s}}}\!\!\sin(\omega\,t + \varphi).
\end{equation}
where  $\mathbf{\Omega}_{s} \subset \mathbb{R}$ is a finite set of frequencies.
The waveforms consist of a fundamental frequency $\Omega$, 
and convolutions of this fundamental with higher order partial frequencies.
Three frequency sets $\mathbf{\Omega}_{s}$  are considered: 
the single fundamental frequency $\mathbf{\Omega}_{1}=\left\{\Omega\right\}$, 
the fundamental and the second partial frequencies
$\mathbf{\Omega}_{2}=\left\{\Omega,2\Omega\right\}$, and
the fundamental, second, and third partial
frequencies $\mathbf{\Omega}_{3}=\left\{\Omega,2\Omega,3\Omega\right\}$.
The fundamental driving frequency $\Omega_f$ is $\Omega$  
for $\mathbf{\Omega}_{1}$ and $\mathbf{\Omega}_{2}$,  
and $2\Omega$ for $\mathbf{\Omega}_{3}$.
The products in Eq.~(\ref{eq:BT}) for the three sets
can be written as finite sums 
\begin{equation}
	\begin{aligned}
    \label{eq:Driving_Sum}
    E_1(t) &= a \sin(\Omega t + \varphi), \\[1ex]
    E_2(t) &= \frac{a}{2}(\cos(\Omega t)+\cos(3 \Omega t + 2 \varphi)), \\[1ex]
		E_3(t) &= \frac{a}{4} (\sin(2\Omega t+\varphi)+ \sin(4\Omega t+\varphi)  \\[1ex]
		 & \quad + \sin(6\Omega t+3\varphi)  + \sin(\varphi)), 	
	\end{aligned}
\end{equation}
where the leading order terms exhibit the characteristic fundamental frequency.
The maximum amplitude of each waveform is set to unity by 
adjusting the value of the parameter $a$ accordingly. 
For the $\mathbf{\Omega}_{1}$, $\mathbf{\Omega}_{2}$, and $\mathbf{\Omega}_{3}$
sets, $a = 1$, $a \approx 1.299$, and $a \approx 1.822$, respectively.
The functional forms of (\ref{eq:Driving_Sum}) are shown in Fig.~\ref{fig:TS_traj}.

\section{\label{sec:TS_traj}The Transition State Trajectory}

%%%%%%%%%%%%%%%%%%%%%
\begin{figure*}
\includegraphics[width=17.0cm,clip]{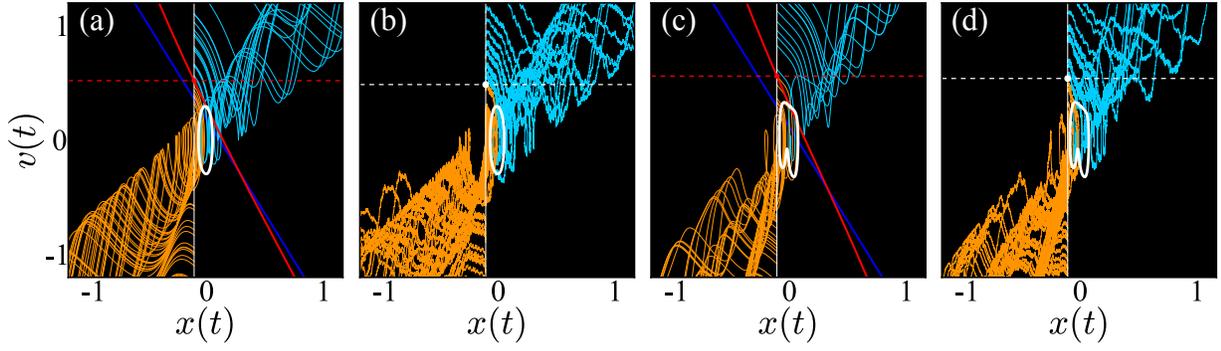}
\caption{\label{fig:Phase}
Phase space plots for a swarm of trajectories following the equations of motion~(\ref{eq:motionAnharm}) 
with various driving frequency sets and parameter values: 
$\mathbf{\Omega}_{1}$ with $\Omega_f=5$ and $\epsilon=1$ for 
(a) $\sigma=0$ (noiseless) and 
(b) $\sigma = 0.025$ (thermal) driving,
$\mathbf{\Omega}_{3}$ with $\Omega_f=4$ and $\epsilon=3$ 
for (c) $\sigma=0$ and 
(d) $\sigma = 0.025$.
The initial position for every trajectory, $x_0=-0.1$, is shown as a vertical solid white line.
Reactive trajectories are colored in cyan and nonreactive trajectories are colored in orange.
The TS~trajectory $\mathbf{\Gamma}^\ddag$ is shown as a solid white curve.
The critical velocity $V^\ddag$ is indicated by a circle 
at the intersection of the dashed horizontal line and the line of initial conditions. 
The solid red curve is the critical curve $V^\ddag_c$ and the solid blue curve is the harmonic critical curve $V^\ddag_c |_{\epsilon=0}$.
Parameters for all panels are $\gamma=1$ and $\varphi=0$ in dimensionless units.}
\end{figure*}
%%%%%%%%%%%%%%%%%%%%%

The construction and existence of a structure whose configuration space 
projection 
is free of recrossings is dependent on the mechanism and geometry of 
a given reaction. 
For example, Mullen \textit{et al.}\cite{Mullen2014,Peters14} have proposed 
that for ion-pair 
dissociation a no-recrossings DS does not exist.
This is in contradiction to earlier work by Truhlar and Garrett\cite{truh2000}
who proposed that through variation of a DS into an optimized orientation, recrossings could be eliminated.
We have previously shown that for a periodically driven system with no thermal driving, 
an optimal (recrossing-free) DS can be readily obtained. It is associated with an unstable PO in the region of 
the BT. \cite{hern14b}
Moreover, a DS that is free of recrossings is known to exist in thermally 
driven systems for the case of a harmonic barrier.\cite{dawn05a,dawn05b}
The time evolution of the configuration space projection of this DS 
has been termed the transition state 
trajectory.\cite{dawn05a,dawn05b,hern06d,Revuelta12,Bartsch12,hern14b,hern14f}
It has not yet been proven is this is the only object 
which is free of recrossings over an arbitrary 
finite time interval.
Nevertheless, all that is needed here is its existence and
the configuration projection of the TS~trajectory defines 
a DS that is recrossing-free.

The TS~trajectory is a specific trajectory that never descends into either the product or reactant regions, 
remaining bounded to BT for all time.  
For the system (\ref{eq:motionAnharm}), it is a moving saddle point 
to which stable and unstable manifolds can be attached.
All trajectories that exponentially approach the TS~trajectory as $t\to\infty$ are contained on the stable manifold.
These trajectories will never descend from the BT region and
therefore separate reactive from nonreactive trajectories in phase space.
The unstable manifold is formed from trajectories that approach the 
TS~trajectory as $t\to-\infty$.
The role of the unstable manifold is less important for the purposes considered here.

For an arbitrary driving $E(t)$ 
of a harmonic ($\epsilon=0$) potential,
the equations of motion
can be solved exactly and an exact form of the TS trajectory can be obtained.
The eigenvalues of (\ref{eq:motionAnharm})
\begin{equation}
	\lambda_{\text s,u} = -\frac{1}{2} \left(\gamma \pm \sqrt{\gamma^2 + 4 \omb^2}\right),
\end{equation}
correspond to the stable and unstable manifolds. 
The $S$ functionals~\cite{dawn05b,Kawai07}
\begin{equation}
    \label{eq:SDef}
    S_\tau[\mu, g;t] = \begin{cases}
            \displaystyle -\int_t^\infty g(\tau)\,\exp(\mu(t-\tau)) \,d\tau \!\!\!
                & :\; \Re\mu>0, \\[3ex]
            \displaystyle +\int_{-\infty}^t g(\tau)\,\exp(\mu(t-\tau)) \,d\tau \!\!\!
                & :\; \Re\mu<0,
        \end{cases}
\end{equation}
obtained as a Green's function solution,
suppress the transient exponential factor in the solution and
return only the equilibrium portion.
In the absence of thermal driving ($\sigma=0$),
the TS~trajectory for a harmonic barrier can therefore be expressed as~\cite{Revuelta12,Bartsch12,hern14b,hern14f}
\begin{equation}
	\begin{aligned}
    \label{eq:TStraj_x}
    x^\ddag(t) &= \frac{\omb^2}{\lu-\ls}\,\left(S[\ls, E;t]-S[\lu, E;t]\right), \\
    v^\ddag(t) &= \frac{\omb^2}{\lu-\ls}\,\left(\ls S[\ls, E;t]-\lu S[\lu, E;t]\right).
	\end{aligned}
\end{equation}

For the case of $T$-periodic motion of a harmonic barrier,
the TS~trajectory can be identified more easily by looking for a bounded solution to the equations of motion. 
For the single frequency $\mathbf{\Omega}_{1}$ case, the ansatz
\begin{equation}
	\begin{aligned}
		\label{eq:TSosc}
		x_1^\ddag(t) &= A \sin(\Omega t + \varphi) + B \cos(\Omega t + \varphi),		
	\end{aligned}
\end{equation}
yields the solution
\begin{equation}
	A  = A_1, \qquad B = B_1,
\end{equation}
where
\begin{equation}
	\begin{aligned}
		\label{eq:TSoscCoeff}
		A_k &= a\,\frac{\omb^2 (\omb^2+(k\Omega)^2)}{(\gamma k \Omega)^2 + (\omb^2+(k\Omega)^2)^2},  \\[0.5ex]
		B_k &= a\,\frac{\omb^2 \gamma k \Omega}{(\gamma k \Omega)^2 + (\omb^2+(k\Omega)^2)^2}.
	\end{aligned}
\end{equation}
In the absence of friction, $\gamma=0$, this simplifies to
\begin{equation}
	A_1= a\,\frac{1}{1+\Omega^2/\omb^2}, \qquad B_1=0.
\end{equation}
In this case, the TS~trajectory will oscillate in phase with the barrier, 
but with smaller amplitude $A_1<a$.

For the $\mathbf{\Omega}_{2}$ and $\mathbf{\Omega}_{3}$ cases, 
ans\"atze can be constructed through Fourier series expansion of Eq.~(\ref{eq:BT}) yielding the solutions
\begin{equation}
 \begin{aligned}
   \label{eq:TSosc2}
   x_2^\ddag(t) &= \tfrac{1}{2}(A_1 \cos(\Omega t)- B_1 \sin(\Omega t) \\[1ex]
   & \quad - A_3 \cos(3\Omega t+2 \varphi) + B_3\sin(3\Omega t+2 \varphi) ), 
 \end{aligned}
\end{equation}
and
%in the $\mathbf{\Omega}_{2}$ case and 
\begin{equation}
	\begin{aligned}
		\label{eq:TSosc3}
		x_3^\ddag(t) &= \tfrac{1}{4}(A_2 \sin(2\Omega t+\varphi)+B_2 \cos(2\Omega t+\varphi) \\[1ex]
		& \quad +A_4 \sin(4\Omega t+\varphi)+ B_4 \cos(4\Omega t+\varphi)  \\[1ex]
		& \quad -A_6 \sin(6\Omega t+3\varphi)-B_6 \cos(6\Omega t+3\varphi)  \\[1ex]
		& \quad +a \sin(\varphi)), 
	\end{aligned}
\end{equation}
respectively.

For periodically driven anharmonic barriers ($\epsilon\ne0$), 
the TS~trajectory rermains an unstable PO\cite{hern14b}
but it does not admit to an exact solution of 
the system of equations (\ref{eq:motionAnharm}).
Nevertheless, the phase space vector of the TS~trajectory 
$\mathbf{\Gamma}^\ddag= (x^\ddag(t),v^\ddag(t))$ is a bounded solution 
to the equations of motion.
To find this bounded solution to arbitrary accuracy, 
numerical Newton-Raphson root finding methods were applied.

The dynamics of $x^\ddag(t)$, shown in Fig.~\ref{fig:TS_traj},
illustrate the result that the instantaneous position of the TS~trajectory 
does not correspond 
to the energetic maximum of the potential surface.
For dissipative systems ($\gamma \neq 0$), 
$x^\ddag(t)$ will either lag behind in phase, 
as is the case for both the $\mathbf{\Omega}_1$ and $\mathbf{\Omega}_3$ sets, or advance in phase as is the case 
for the $\mathbf{\Omega}_2$ set, with respect to motion defined by $E(t)$. 
Also note that $x^\ddag(t)$ oscillates with a smaller amplitude than $E(t)$.
Thus, even for in-phase oscillations, e.g., when $\gamma=0$, 
it will not correspond to the location of an energetic saddle point.
Figure~\ref{fig:TS_traj} also shows the dependence of $x^\ddag(t)$ 
on the anharmonic parameter $\epsilon$. 
As $\epsilon$ is increased, the curvature of the energy barrier increases. 
Non-intuitively, this results in a larger amplitude of oscillation for 
$x^\ddag(t)$ to remain bounded to the BT. 
This trend persists for all $\mathbf{\Omega}_s$. 

For dynamical analysis, it is advantageous to introduce 
a coordinate system which has a fixed point at the origin. 
In relative coordinates
\begin{equation}
	\Delta x = x - x^\ddag(t), \qquad \Delta v = v - v^\ddag(t),
\end{equation}
the equations of motion read
\begin{equation}
	\begin{aligned}
		\label{eq:motionRel}
		\Delta \dot x &= \Delta v, \\
		\Delta \dot v &= -\gamma \Delta v -U'(\Delta x + x^\ddag(t)) + U'(x^\ddag(t)).
	\end{aligned}
\end{equation}
The relative equations of motion have a fixed point $\Delta \mathbf{\Gamma}^\star$ 
at $\Delta x = \Delta v = 0$, i.e., on the TS~trajectory, 
and the surrounding vector field itself will now oscillate with period $T$, 
the same period as the driving.
The TS~trajectory has both a stable and an unstable manifold attached.  
In relative coordinates, the directions of these manifolds will depend on time.

\section{\label{sec:Rate_Theory}Reaction Rate Theory}
In the TST formalism, 
the rate of a chemical reaction is given by the time-dependence of the conversion process from 
reactant to product ($\text{R} \rightarrow \text{P}$) 
where a DS in either configuration space or phase space
separates the reactive constituents.
The reaction rate can be obtained from the dynamics of the normalized reactive population 
($P_\text{R} \rightarrow P_\text{P}$) either through analytical propagation of the phase space density of initial conditions 
or by treating large numbers of trajectories as discrete sets, and integrating the equations of motion.

Consider a set of trajectories evolving through (\ref{eq:motionAnharm}) 
that all have initial positions $x_0<x^\ddag(0)$ on the reactant side of the moving surface. 
The initial position distribution at time $t=0$ is $\delta(x-x_0)$ and the  
initial phase space density is
\begin{equation}
	\label{eq:density_0_posvel}
	p_0( x,v) = \delta( x -  x_0)\, q(v)
\end{equation}
where $q(v)$ is a Boltzmann distribution. 
The initial velocity $v_0$ of each trajectory is sampled from $q(v)$, 
although at later times due to dissipation and driving this distribution will not be conserved.
A fraction of this initial density contains reactive trajectories. 
From the survival probability of $P_\text{R}$ the reaction rate 
can be expressed as the instantaneous flux-over-population.
The flux calculation is formally exact because the DS attached 
to the TS~trajectory is recrossing-free.

\subsection{\label{harTS}Harmonic barriers}
When the barrier is harmonic $(\epsilon=0)$, reactive trajectories will cross the moving DS at a time \cite{hern06d}
\begin{equation}
	\label{eq:tCross}
	t^\ddag = \frac{1}{\lu-\ls} \ln \left(\frac{\Delta v_0 - \lu\Delta x_0}{\Delta v_0 - \ls \Delta x_0}\right).		
\end{equation}
The crossing time is a monotonically decreasing function of the initial velocity $\Delta v_0$: fast trajectories cross earlier. It diverges as $\Delta v_0\to \ls\Delta x_0$ approaches the stable manifold, and it tends to zero as $\Delta v_0\to\infty$.

At any time $t>0$, the product region $\Delta x>0$, 
to the right of the moving surface, will contain all those trajectories that cross the surface at a time $t^\ddag<t$.
These are the trajectories that have an initial velocity of at least 
$\vmin = v^\ddag(0) + \Delta \vmin$, where $t^\ddag(\Delta\vmin) = t$. 
From this condition, we obtain
\begin{equation}
	\label{eq:vmin}
	\Delta\vmin = \frac{\lu e^{-\lu t} - \ls e^{-\ls t}}{e^{-\lu t} - e^{-\ls t}}\,\Delta x_0.
\end{equation}
The population of the product region at time $t$ is therefore
\begin{equation}
	\label{eq:pPop}
	P_\text{P}(t) = \int_{\vmin(t)}^\infty q(v)\,dv,
\end{equation}
and the flux across the moving surface is
\begin{equation}
	\begin{aligned}
		\label{eq:popFlux}
			F_\text{M}(t) &= \frac{dP_\text{P}}{dt}  \\
			&= -q(\vmin(t))\, \frac{d\vmin}{dt} \\
			&= -q(\vmin(t))\, \frac{d\Delta\vmin}{dt}  \\
			&= -q(\vmin(t))\, \Delta x_0\, (\lu-\ls)^2\,\frac{e^{(\lu+\ls)t}}{(e^{\lu t}-e^{\ls t})^2}.
	\end{aligned}
\end{equation}
This result is positive because $\Delta x_0<0$.

Alternatively, the flux can be calculated directly from the flux integral
\begin{equation}
	\label{eq:fluxInt}
	F_\text{M}(t) = \int_0^\infty d\Delta v\,\Delta v\,p_t(\Delta x=0,\Delta v),
\end{equation}
where $p_t(\Delta x,\Delta v)$ is the density of trajectories in phase space at time $t$. 
Initially, this density is
\begin{equation}
	\label{eq:density_0}
	p_0(\Delta x,\Delta v) = \delta(\Delta x - \Delta x_0)\, q(v^\ddag(0)+\Delta v).
\end{equation}
At later times, it can be obtained from
\begin{equation}
	\label{eq:density_t}
	p_t(\Delta x,\Delta v) = e^{\gamma t}\, p_0(\Delta x(-t),\Delta v(-t)).
\end{equation}
Here $\Delta x(-t)$ and $\Delta v(-t)$ denote the phase space point reached from $\Delta x, 
\Delta v$ by propagating backwards to $-t$, i.e., 
it is the initial condition that has reached $\Delta x, \Delta v$ at time $t$. 
The exponential prefactor accounts for the shrinkage of phase space volume: 
The relative dynamics stretches distances at a rate $\lu$ in the $u$ direction and by a rate $\ls<0$ in the $s$ direction.
 Volumes therefore are ``stretched'' at a constant rate $\lu+\ls=-\gamma<0$, and densities must increase accordingly.

Because the relative dynamics is linear, the equations of motion can be solved explicitly. The result is
\begin{align*}
	\Delta x(-t) &= a_x\,\Delta x + a_v \,\Delta v, \\
	\Delta v(-t) &= b_x\,\Delta x + b_v\,\Delta v,
\end{align*}
with
\begin{align*}
	a_x &= \frac{\lu \,e^{-\ls t} - \ls \,e^{-\lu t}}{\lu-\ls}, &
	a_v &= \frac{e^{-\lu t} - e^{-\ls t}}{\lu-\ls}<0,  \\[1ex]
	b_x &= -\frac{\lu\ls (e^{-\lu t} - e^{-\ls t})}{\lu-\ls}, &
	b_v &= \frac{\lu\,e^{-\lu t} -\ls\, e^{-\ls t}}{\lu-\ls}.	
\end{align*}
We thus obtain the flux integral
\begin{widetext}
\begin{equation}
	\begin{aligned}
		\label{eq:intFlux}
			F_\text{M}(t) &= e^{\gamma t} \, \int_0^\infty d\Delta v\, \Delta v\, \delta(a_v\,\Delta v-x_0)\,
					q(v^\ddag(0)+b_v\,\Delta v)  \\[1ex]
			&= e^{\gamma t} \, \int_0^\infty d\Delta v\, \Delta v\, \frac{\delta\left(\Delta v-x_0/a_v\right)}{|a_v|}\,
					q(v^\ddag(0)+b_v\,\Delta v)  \\[1ex]
			&= \frac{e^{\gamma t}}{-a_v}\, \frac{x_0}{a_v}\, q\left(v^\ddag(0) + b_v/a_v\,x_0\right),
	\end{aligned}
\end{equation}
\end{widetext}
which can be shown to agree with Eq.~\eqref{eq:popFlux}.

In the limit $t\to\infty$, the minimum velocity~\eqref{eq:vmin} is approximately
\begin{eqnarray}
	\Delta\vmin &=& \ls\,\Delta x_0 - (\lu-\ls)\Delta x_0\,e^{-(\lu-\ls)t}\nonumber \\[1ex]
	&  &+ {\cal O}\left(e^{-2(\lu-\ls)t}\right).
\end{eqnarray}
As expected, it tends to $\ls\,\Delta x_0$, which is the location of the stable manifold. Therefore,
\begin{equation}
	\vmin(\infty) = v^\ddag(0) + \ls\,\Delta x_0 = V^\ddag. 
\end{equation}
The critical velocity $V^\ddag$ is determined by the point of intersection
between the stable manifold and the line $x=x_0$ of initial conditions.\cite{Revuelta12,Bartsch12}.
The identification of $V^\ddag$ allows the separation of reactive ($v_0>V^\ddag$) and nonreactive trajectories ($v_0<V^\ddag$) from initial conditions.
The stable manifold at $t = 0$ can be calculated through extension of this point to all values of $x_0$
and defines a critical curve $V^\ddag_c$.
As illustrated in Figs.~\ref{fig:Phase}(a) and \ref{fig:Phase}(c),  
$V^\ddag_c$ is a time-invariant phase space object 
which separates the reactive and nonreactive basins.

%%%%%%%%%%%%%%%%%%%%%
\begin{figure}
\includegraphics[width=8.5cm,clip]{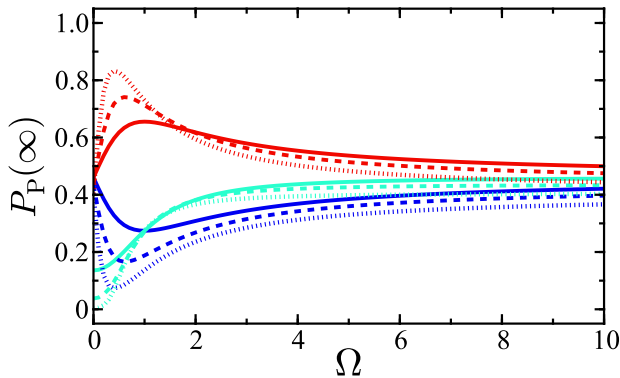}
\caption{\label{fig:Pinf}
The asymptotic product population $P_\text{P}(\infty)$ of 
the harmonic potential ($\epsilon = 0$) as a function of 
driving frequency $\Omega$ for the $\mathbf{\Omega}_{1}$ frequency set. 
The curves are colored with respect to 
the value of the initial phase shift: 
$\varphi=0$ (blue), $\varphi=\pi/2$ (cyan), and $\varphi=\pi$ (red). 
For each value of $\varphi$, the dependency of the asymptotic population 
on the friction parameter $\gamma$ is shown by varying the linestyle:
 $\gamma=0$ (solid), $\gamma=1$ (dashed), and $\gamma=2$ (dotted). 
}
\end{figure}
%%%%%%%%%%%%%%%%%%%%%

The product population in the long-time limit is
\begin{equation}
	P_\text{P}(\infty) = \int_{v^\ddag(0)+\ls\,\Delta x_0}^\infty q(v)\,dv.
\end{equation}
As shown in Fig.~\ref{fig:Pinf}, 
the asymptotic population of the product region depends strongly on the frequency of the barrier motion $\Omega$, 
the initial phase $\varphi$, and the friction $\gamma$.
The asymptotic value is approached according to 
\begin{eqnarray}
	\label{eq:fluxAsymptHarm}
	P_\text{P}(t) &=& P_\text{P}(\infty) - \int_{\vmin(\infty)}^{\vmin(t)} q(v)\,dv \nonumber \\[1ex]
		&=& P_\text{P}(\infty) + q(\vmin(\infty))\,(\lu-\ls)\,\Delta x_0\,e^{-(\lu-\ls)t} \nonumber \\[1ex]
		&  &+{\cal O}\left(e^{-2(\lu-\ls)t}\right).
\end{eqnarray}
The rate of approach, i.e., the barrier crossing rate is 
\begin{equation}
	\label{eq:ratehar}
	\lu-\ls = \sqrt{\gamma^2+4\omb^2}.
\end{equation} 
It depends only on the damping and the shape of the barrier, 
but not on the details of the barrier motion or the distribution of initial conditions (unless $q(\vmin(\infty))$ happens to vanish).

\subsection{\label{anharTS}Anharmonic barriers}

Analogous to the harmonic case, 
for an anharmonic barrier,
we assume that there is a unstable PO with the period of the driving. 
This is similar to the POs used by Lehmann \textit{et al}.\cite{Lehmann00a,Lehmann00b,Lehmann03} 
for the case of thermal activation with additive periodic driving.
Note that this TS~trajectory  $\mathbf{\Gamma}^\ddag$ is an exact solution to the equations of motion.
As $\mathbf{\Gamma}^\ddag$ is an unstable PO, it has stable and unstable manifolds attached. 
The manifolds are uniquely defined and can be calculated perturbatively \cite{Revuelta12,Bartsch12} or with a numerical scheme. 
The dependence of these manifolds on $\epsilon$ 
and the corresponding phase space reaction dynamics are shown in Figs.~\ref{fig:Phase}(a) and \ref{fig:Phase}(c). 
With increasing anharmonicity, $V^\ddag$ also increases due to curvature in the stable manifold.
This results in a decrease in fraction of trajectories that 
surmount the barrier leading to products.

The nonlinear equations of motion~\eqref{eq:motionRel} cannot, in general, be solved exactly. Let 
\begin{equation}
	\mathbf{\Phi} (\mathbf{\Gamma}_{\!0},t_0; t) = 
		\begin{pmatrix}
			\varphi_x (\mathbf{\Gamma}_{\!0},t_0; t) \\
			\varphi_v (\mathbf{\Gamma}_{\!0},t_0; t)
		\end{pmatrix}
\end{equation}
represent the phase space point that is reached at time $t$ by a trajectory that starts at $\mathbf{\Gamma}_{\!0}$ at time $t_0$. Because of the external driving, 
it depends on $t$ and $t_0$ separately, 
not only on the difference $t-t_0$. 
The Jacobian matrix of this trajectory with respect to the initial conditions is
\begin{equation}
	\label{eq:JDef}
	\arraycolsep=4.0pt
	\setlength{\delimitershortfall}{-1pt}
	\boldsymbol{J}(\mathbf{\Gamma}_{\!0},t_0;t) = \begin{pmatrix}
		\dfrac{\partial \varphi_x}{\partial x_0} & \dfrac{\partial \varphi_x }{\partial v_0} \\[2.3ex]
		\dfrac{\partial \varphi_v} {\partial x_0} & \dfrac{\partial \varphi_v }{ \partial v_0}
	\end{pmatrix}.
\end{equation}
All derivatives on the right hand side of~\eqref{eq:JDef} are to be evaluated at $(\mathbf{\Gamma}_{\!0},t_0;t)$.

Reactive trajectories are those that have an initial velocity $v_i > V^\ddag$, 
for a critical velocity $V^\ddag$ measured at $x_0$. 
Each reactive trajectory will cross the moving DS at a time $t_c(\Delta v_i)$ and with a velocity $v_c$. 
If the crossing time decays monotonically from 
$t_c(\Delta V^\ddag)=\infty$ to $t_c(\infty)=0$ 
the inverse function $\Delta v_i(t_c)$ or $v_i(t_c) = v^\ddag(0) + \delta v_i(t_c)$ can be obtained. 
For any crossing time $t_c>0$, there is a unique initial velocity $v_i$ that will lead to a crossing at the given time.

The population of the product region $\Delta x>0$ at time $t_f$ is therefore, as in Eq.~\eqref{eq:pPop},
\begin{equation}
	P_\text{P}(t_c) = \int_{v_i(t_c)}^\infty q(v)\,dv,
\end{equation}
and the flux across the moving surface is
	\begin{align}
		\label{eq:popFluxGen}
		F_\text{M}(t_c) &= \frac{dP_\text{P}}{dt_c} \nonumber \\
			&= -q(v_i(t_c))\, \frac{d v_i}{dt_c} \nonumber \\[1ex]
			&= -q(v_i(t_c))\, \frac{d\Delta v_i}{dt_c}
	\end{align}
This result is positive because the initial velocity is a decreasing function of the crossing time.

The flux can also be evaluated directly from the flux integral~(\ref{eq:fluxInt})
\begin{equation}
	F_\text{M}(t_c) = \int_0^\infty d\Delta v\,\Delta v\,p_{t_c}(\Delta x=0,\Delta v),
\end{equation}
where $p_t(\Delta x,\Delta v)$ is the density of trajectories in phase space at time $t$. 
Initially, this density is (Eq.~\eqref{eq:density_0})
\begin{equation}
	p_0(\Delta x,\Delta v) = \delta(\Delta x - \Delta x_0)\, q(v^\ddag(0)+\Delta v).
\end{equation}
At later times, it can be obtained from Eq.~\eqref{eq:density_t}
\begin{equation}
	p_t(\Delta x,\Delta v) = e^{\gamma t}\, p_0(\varphi_x(\Delta x,\Delta v,t;0), \varphi_v(\Delta x,\Delta v,t;0)).
\end{equation}
Here we have used the general notation for the flow of the equation of motion. 
The exponential accounts for the shrinkage of phase space volume and the corresponding increase in density.
 It is the same as in the harmonic case: In general, the flow of a differential equation 
$\dot{\boldsymbol{u}}=\boldsymbol{f}(\boldsymbol{u})$ 
leads to a stretching of volume whose rate is the divergence of the vector field $\boldsymbol{f}$. 
For Eq.~\eqref{eq:motionRel}, this rate is constant $-\gamma$, 
so that over time $t$ all volumes will shrink by a factor $e^{-\gamma t}$. 

The flux integral formula~\eqref{eq:fluxInt} now reads
\begin{widetext}
	\begin{equation}
		F_\text{M}(t_c) = e^{\gamma t_c} \int_0^\infty d\Delta v\, \Delta v\, 
			\delta(\varphi_x(0,\Delta v,t_c; 0)-\Delta x_0)\;
			q(v^\ddag(0) + \varphi_v(0,\Delta v,t_c; 0)).
	\end{equation}
\end{widetext}
The $\delta$~function requires that the trajectory that reaches $\Delta x=0, \Delta v$ at time $t_c$ must have started at $\Delta x_0$ at time 0. 
It produces a single contribution to the integral at velocity $\Delta v_c(t_c)$, so that
\begin{equation}
	\label{eq:intFlux2}
	F_\text{M}(t_c) = e^{\gamma t_c}\,q(v^\ddag(0)+\Delta v_i(t_c))\,
		\frac{\Delta v_c(t_c)}{\left|\left.\frac{\partial \varphi_x}{\partial \Delta v_0}\right|_{t_c}\right|},
\end{equation}
where $\varphi_v(0,\Delta v,t_c; 0) = \Delta v_i(t_c)$ and the subscript $t_c$ indicates that the derivative is to be evaluated at $(0,\Delta v_c(t_c),t_c;0)$. 
Similarly, a subscript $0$ will be used to require evaluation at $(\Delta x_0,\Delta v_i(t_c),0; t_c)$. 
These subscripts indicate derivatives of the flow taken along the trajectory from $(\Delta x_0, \Delta v_i(t_c))$ at $t=0$ 
forward in time to $(0,\Delta v_c(t_c))$ at $t=t_c$ (subscript 0) and along the same trajectory backward in time (subscript $t_c$).

To verify that the flux integral~\eqref{eq:intFlux2} gives the same result as~\eqref{eq:popFluxGen} that was obtained from the product population, it must be shown that
\begin{equation}
	\label{eq:fluxCond1}
	-\frac{d\Delta v_i}{d t_c} = 
		e^{\gamma t_c}\,
		\frac{\Delta v_c(t_c)}{\left|\left.\frac{\partial \varphi_x}{\partial \Delta v_0}\right|_{t_c}\right|}.
\end{equation}
To this end, first note that $\Delta v_i(t_c)$ is defined by the condition
\[
	\varphi_x(\Delta x_0, \Delta v_i(t_c), 0; t_c)=0.
\]
Differentiating this condition with respect to $t_c$ gives,
\begin{equation}
	\label{eq:derivCond}
	\left.\frac{\partial \varphi_x}{\partial \Delta v}\right|_0 \, \frac{d\Delta v_i}{dt_c}
	+ \left.\frac{\partial \varphi_x}{\partial t}\right|_0
	= 0.
\end{equation}
Now $\partial \varphi_x/\partial t$ is the velocity of the trajectory at the end point. 
The second term in Eq.~\eqref{eq:derivCond} is therefore $\Delta v_c(t_c)$. 
With this result, the condition~\eqref{eq:fluxCond1} simplifies to
\begin{equation}
	\label{eq:fluxCond2}
	-\left.\frac{\partial \varphi_x}{\partial \Delta v}\right|_{t_c} = e^{\gamma t_c}\,
	\left.\frac{\partial \varphi_x}{\partial \Delta v}\right|_0.
\end{equation}
Under the given assumptions on the geometry, the derivative on the left hand side is negative: 
A trajectory that arrives at the DS with larger velocity must have started further away, i.e., at smaller $\Delta x(0)$.

The derivatives occurring in Eq.~\eqref{eq:fluxCond2} are elements of the Jacobian matrices 
\begin{equation}
	\arraycolsep=3.1pt
	\setlength{\delimitershortfall}{-1pt}
	\boldsymbol{J}|_{t_c}=\boldsymbol{J}(0, \Delta v_c(t_c), t_c; 0)
	 = \begin{pmatrix}
		\dfrac{\partial \varphi_x}{  \partial x}\Big|_{t_c} & \dfrac{\partial \varphi_x }{ \partial v}\Big|_{t_c} \\[2.3ex]
		\dfrac{\partial \varphi_v}{  \partial x}\Big|_{t_c} & \dfrac{\partial \varphi_v }{ \partial v}\Big|_{t_c}
	 \end{pmatrix} \nonumber
\end{equation}
and
\begin{equation}
	\arraycolsep=4.5pt
	\setlength{\delimitershortfall}{-1pt}
	\boldsymbol{J}|_0=\boldsymbol{J}(\Delta x_0, \Delta v_i(t_c), 0; t_c)
	 = \begin{pmatrix}
		\dfrac{\partial \varphi_x}{  \partial x}\Big|_{0} & \dfrac{\partial \varphi_x }{ \partial v}\Big|_{0} \\[2.3ex]
		\dfrac{\partial \varphi_v}{  \partial x}\Big|_{0} & \dfrac{\partial \varphi_v }{ \partial v}\Big|_{0}
	 \end{pmatrix} \nonumber
\end{equation}
respectively. Because these matrices describe variations around the same trajectory, 
taken forwards and backwards in time, 
they are inverse to each other. 
Formally, this can be shown by taking derivatives of the flow property
\[
	\mathbf{\Phi}(\mathbf{\Phi}(\mathbf{\Gamma}_{\!0},0;t_c),t_c;0) = \mathbf{\Gamma}_{\!0} \qquad
	\text{for all $\mathbf{\Gamma}_{\!0}$},
\]
which says that propagating an arbitrary phase space point $\mathbf{\Gamma}_{\!0}$ 
forward in time by $t_c$ and back again will return the original point.

By the well known formula for the inverse of a $2\times 2$ matrix, it follows that
\begin{equation}
	\arraycolsep=4.5pt
	\setlength{\delimitershortfall}{-1pt}
	\begin{aligned}
		\boldsymbol{J}|_{t_c} &= (\boldsymbol{J}|_0)^{-1} \nonumber \\
			&= \frac 1{\det \boldsymbol{J}|_0}
				\begin{pmatrix}[r]
					\dfrac{\partial \varphi_v }{ \partial v}\Big|_{0}& - \dfrac{\partial \varphi_x }{ \partial v}\Big|_{0} \\[2.3ex]
					- \dfrac{\partial \varphi_v }{ \partial x}\Big|_{0}& \dfrac{\partial \varphi_x } {\partial x}\Big|_{0}
				\end{pmatrix},
	\end{aligned}
\end{equation}
so that
\[
	\left.\frac{\partial \varphi_x}{\partial \Delta v}\right|_{t_c}
	 = -\frac{1}{\det \boldsymbol{J}|_0} \left.\frac{\partial \varphi_x}{\partial \Delta v}\right|_0.
\]
Now
\[
	\det \boldsymbol{J}|_0 = e^{-\gamma t_c}
\]
is the factor by which phase space volumes shrink during time $t_c$. 
This proves the condition~\eqref{eq:fluxCond2} and therefore the equality of the two flux formulas.

\subsection{\label{sec:Floquet}Dynamics near the TS}
The TS~trajectory is a moving saddle point and thus
trajectories in the neighborhood of $\mathbf{\Gamma}^\ddag$
can be described by a linearization of the equations of motion.
In the phase space vector relative coordinate 
$\Delta \mathbf{\Gamma}=(\Delta x,\Delta v)$ this linearization is given by 
\begin{equation}
	\label{eq:gamma_sys}
	\Delta\dot{\mathbf{\Gamma}} =\boldsymbol{J}(t)\,\Delta\mathbf{\Gamma}
\end{equation}
where
\begin{equation}
	\label{eq:lin_Jac}
	\arraycolsep=5pt\def\arraystretch{1.5}
		\boldsymbol{J}(t)=
		\begin{pmatrix}
			0 & 1 \\
			\omb^2 +3 \epsilon(x^\ddag(t)-E(t))^2  & -\gamma
		\end{pmatrix}
\end{equation}
is the Jacobian of Eq.~(\ref{eq:motionRel}).
The asymptotic decay rate of $P_\text{R}(t)$ is determined by 
the behavior of trajectories with initial conditions close to the stable manifold.
For an ensemble of trajectories constituting an initial phase space density $p_0$, 
trajectories that emanate close to $V_c^\ddag$ (the stable manifold at $t=0$) 
will persist in the neighborhood 
where \eqref{eq:gamma_sys} is valid for long times. 
The decay of these trajectories determines the reaction rate.

The stretching and compression of phase space about a PO is known to dictate escape rates
in conservative\cite{kadanoff84,skodje90,gaspard98} and dissipative systems. \cite{hern14f}
When $\boldsymbol{J}(t)$ is periodic in 
systems of the form of Eq.~\eqref{eq:gamma_sys},
the rate of deformation in the linearized phase space can be quantified 
through calculation of the Floquet exponents.\cite{Skokos01}

To classify the stability of $\Delta \mathbf{\Gamma}^\ddag$ we consider the dynamics of a perturbation vector $\Delta \boldsymbol{\sigma}(t)$. 
The equation of motion (\ref{eq:gamma_sys}) is linear in $\Delta \boldsymbol{\sigma}(t)$ and thus it satisfies
\begin{equation}
	\label{eq:ep_sys}
	\Delta\dot{\boldsymbol{\sigma}} =\boldsymbol{J}(t)\,\Delta\boldsymbol{\sigma},\;\;\; \Delta\boldsymbol{\sigma}(0)=\boldsymbol{I},
\end{equation}
where $\boldsymbol{I}$ is the $2 \times 2$ identity matrix.
The principal fundamental matrix solution over one period of the driving is the monodromy matrix
\begin{equation}
	\label{eq:Mono}
	\arraycolsep=5pt\def\arraystretch{1.5}
		\boldsymbol{M}=
		\begin{pmatrix}
			\Delta\sigma^{(1)}(T)& \Delta\sigma^{(2)}(T) \\
			\Delta\dot{\sigma}^{(1)}(T)  &\Delta\dot{\sigma}^{(2)}(T)
		\end{pmatrix}.
\end{equation}
A fundamental matrix solution $\Delta\boldsymbol{\sigma}(t)$ of \eqref{eq:gamma_sys} at some later time $t+ k T$, 
for $k = 1,2,3\ldots$, can be obtained as
\begin{equation}
\Delta\boldsymbol{\sigma}(t+ k T) = \boldsymbol{M}^k\Delta\boldsymbol{\sigma}(t),
\end{equation}
through repeated operation by the monodromy matrix.

The eigenvalues $m_{\text s,u}$  of $\boldsymbol{M}$ are the Floquet multipliers.
The Floquet exponents
\begin{equation}
 \mu_{\text s,u} =\frac{1}{T} \ln|m_{\text s,u}|
\end{equation}
quantify the stability of  $\Delta \mathbf{\Gamma}^\ddag$ 
and give the rate of expansion or contraction of the perturbation of per unit time. \cite{chaosbook,Boland09,Flo13}
The TS~trajectory has both an unstable $\mu_u>0$ and a stable $\mu_s<0$ exponent which correspond to 
stretching and contraction of the initial perturbation in the directions of the unstable and stable manifolds, respectively. 
 
For an arbitrary time interval of length $T$, trajectories that cross the DS in this interval form a strip in the phase plane. 
Trajectories that cross the DS in the next following time interval $T$ form a similar strip that that is closer to the stable manifold.
In the region where the linearized system is valid, the phase space density is constant.
The flux of trajectories through the DS in a given time interval is proportional to the width of the strip that contains these trajectories. 
During sequential periods this width decreases by a factor $e^{-(\mu_u-\mu_s)t}$. 
From this it follows that, up to periodic modulation, the flux must decay as $e^{-(\mu_u-\mu_s)t}$
and the barrier crossing rate is
\begin{equation}
	\label{eq:rateanhar}
	k_\mathrm{f} = \mu_u-\mu_s,
\end{equation}
which expresses the reaction rate in terms of the characteristic Floquet exponents of the TS~trajectory.
Equation~(\ref{eq:rateanhar}) generalizes Eq.~(\ref{eq:ratehar}) for the case of an anharmonic barrier.

\section{\label{sec:Rates}Numerical results and comparison with theory}

%%%%%%%%%%%%%%%%%%%%%
\begin{figure}
\includegraphics[width = 8.5cm,clip]{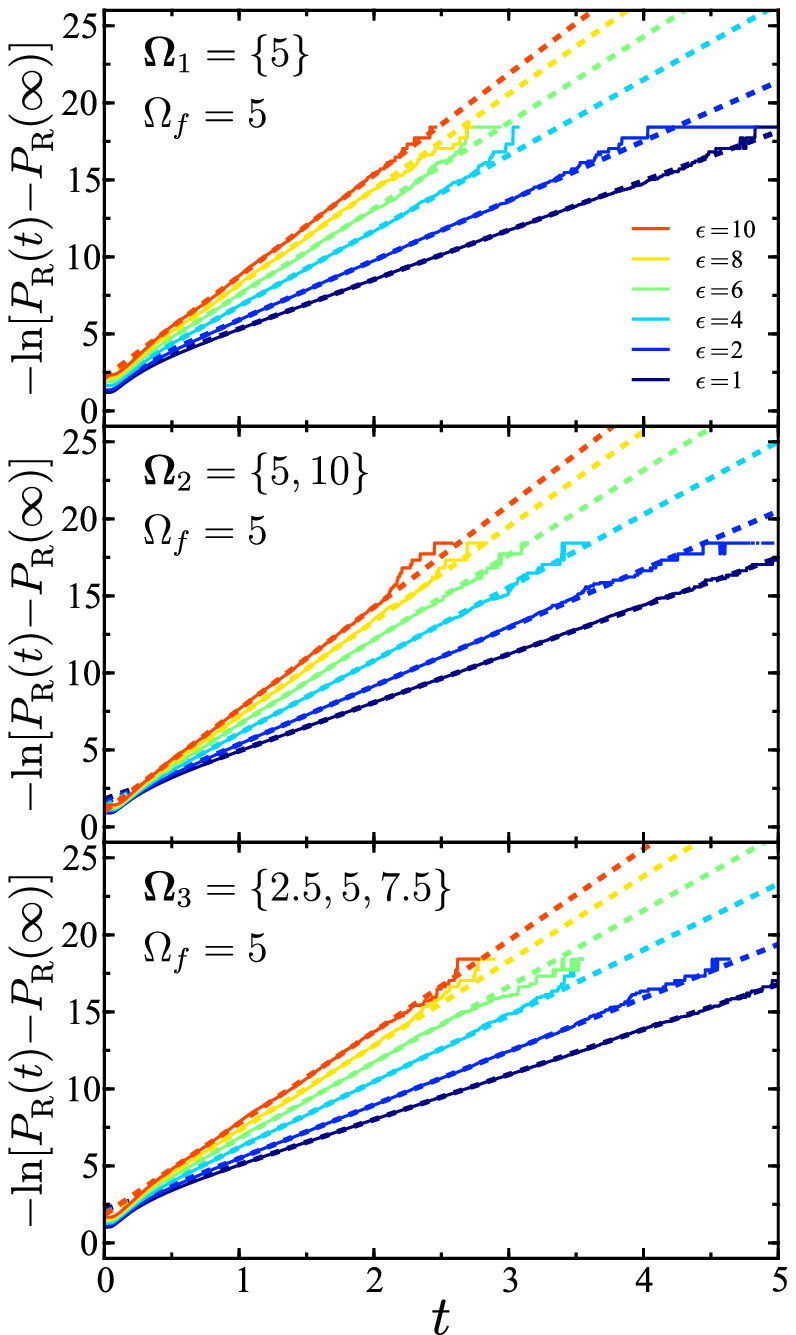}
\caption{\label{fig:Rates}
Time dependence of the scaled logarithm of the reactant population, $-\ln{\left[P_\text{R}(t)-P_\text{R}(\infty)\right]}$, 
for $\mathbf{\Omega}_1$ (top), $\mathbf{\Omega}_2$ (middle), and $\mathbf{\Omega}_3$ (bottom) with $\Omega_f=5$ for all panels.
Values of the anharmonic parameter are
$\epsilon\in{\left\{1,2,4,6,8,10\right\}}$.
The slope of each dashed line is the barrier crossing rate $k_\mathrm{f}$.
The color of each line corresponds to the respective $\epsilon$ value.
In all panels, parameters are 
$\gamma=1$ and $\varphi=0$.}
\end{figure}
%%%%%%%%%%%%%%%%%%%%% 

The reaction rate of \eqref{eq:motionAnharm} was calculated by 
simulating ensembles of $n = 10^8-10^9$
trajectories for various sets of parameters 
$\left\{\Omega,\gamma,\epsilon, \sigma \right\}$ 
and following the survival probability of $P_\text{R}$ as a function of time. 
A Runge-Kutta-Maruyama scheme \cite{NaessandMoe2000} was implemented to perform the integration.
In the absence of noise $(\sigma=0)$, this algorithm is the well-known fourth-order Runga-Kutta method.
For all numerical simulations non-dimensional
parameters were used by choosing units such that the barrier frequency $\omb$ and driving amplitude are unity.
Each trajectory was given an initial position $x_0=-0.1$ (in the reactant region) 
and $v_0$ was sampled from a Boltzmann distribution with $k_\text{B} T = 1$.
The choice of initial conditions is arbitrary as the asymptotic decay rate of $P_\text{R}(t)$ is independent of the choice of initial distribution, 
suffice that there is enough density about the stable manifold such that a rate exists. \cite{hern14f}

%%%%%%%%%%%%%%%%%%%%%
\begin{figure*}
\includegraphics[width = 17.0cm,clip]{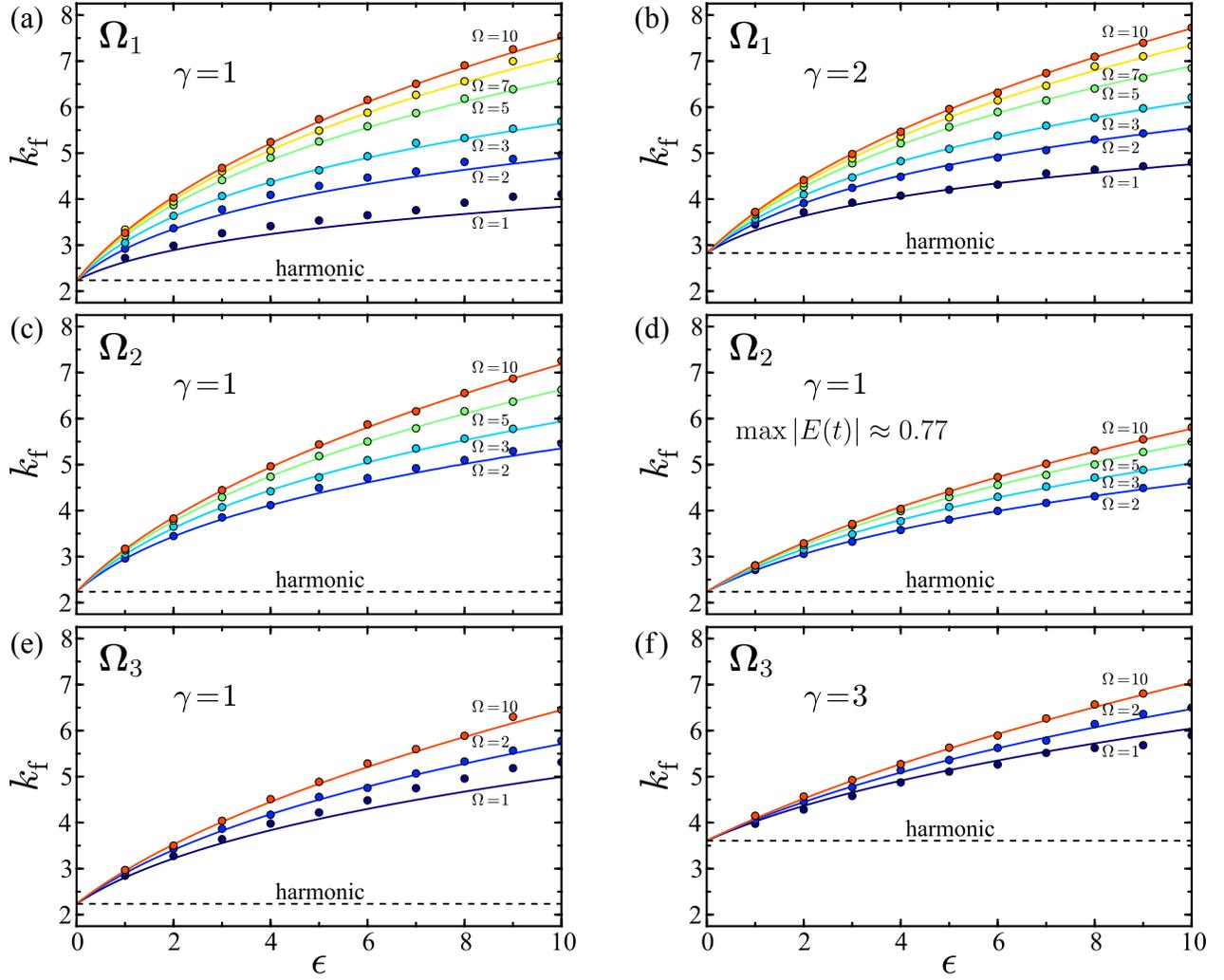}
\caption{\label{fig:Flo}
The barrier crossing rates of systems following the equations of motion~(\ref{eq:motionAnharm}) as a function of
the anharmonic parameter $\epsilon$ for various frequency sets $\mathbf{\Omega}_s$, driving frequencies $\Omega$,
and values of friction $\gamma$, as denoted in each panel.
The circles denote the rates $k_\mathrm{f}$ calculated from the 
time evolution of $P_\text{R}(t)$ through numerically simulation and  
correspond to the dashed lines in Fig. \ref{fig:Rates}.
The solid curves are the rates predicted by the difference in 
the characteristic Floquet exponents $\mu_\text{u}-\mu_\text{s}$ of the corresponding TS~trajectory. 
}
\end{figure*}
%%%%%%%%%%%%%%%%%%%%%

The ensemble of $n$ trajectories was evolved through the equations of motion (\ref{eq:motionAnharm}). 
The normalized reactant population was calculated at each time step in the integration scheme.
An indicator function was employed to follow the state evolution of each trajectory, 
\begin{equation}
     \label{eq:indicator}
 h_\text{R}[x(t)] = \left\{
     \begin{array}{lc}
       0, &  x(t)>x^\ddag(t) \;,\\
			 1, &  x(t)<x^\ddag(t) \;,
     \end{array}
   \right.
\end{equation}
where $x^\ddag(t)$ is the configuration space projection of the TS~trajectory.
If for a specific trajectory $i$, $x_i(t)>x^\ddag(t)$ 
that trajectory is in the product state and is not counted in the reactant population at time $t$. 
The instantaneous normalized population of the reactant region can be found 
by summing over all $n$ trajectories and then normalizing by a factor $1/n$,
\begin{equation}
	\label{eq:population}
	P_\text{R}(t) = \frac{1}{n}\,\sum_{i=1}^n h_\text{R}[x_i(t)].
\end{equation}
Trajectories can only exist in one of two states, 
reactant or product, and so the normalized product population $P_\text{P}=1-P_\text{R}$.

As shown in Fig.~\ref{fig:Rates}, the scaled logarithm of 
the normalized reactant population, $-\ln{\left[P_\text{R}(t)-P_\text{R}(\infty)\right]}$, is approximately linear in time
after an initial transient section implying a first-order rate process.
The asymptotic reaction rate $k_\mathrm{f}$
can thus be found as the slope of the scaled logarithmic curve in the long-time limit.
Periodic modulation in the decay of $P_\text{R}(t)$ was found to become more prominent for low frequency driving $(\Omega_f \lessapprox 2)$.
In these cases the global exponential rate was calculated as an average over these modulations.

%%%%%%%%%%%%%%%%%%%%%
\begin{figure*}
\includegraphics[width = 17.0cm,clip]{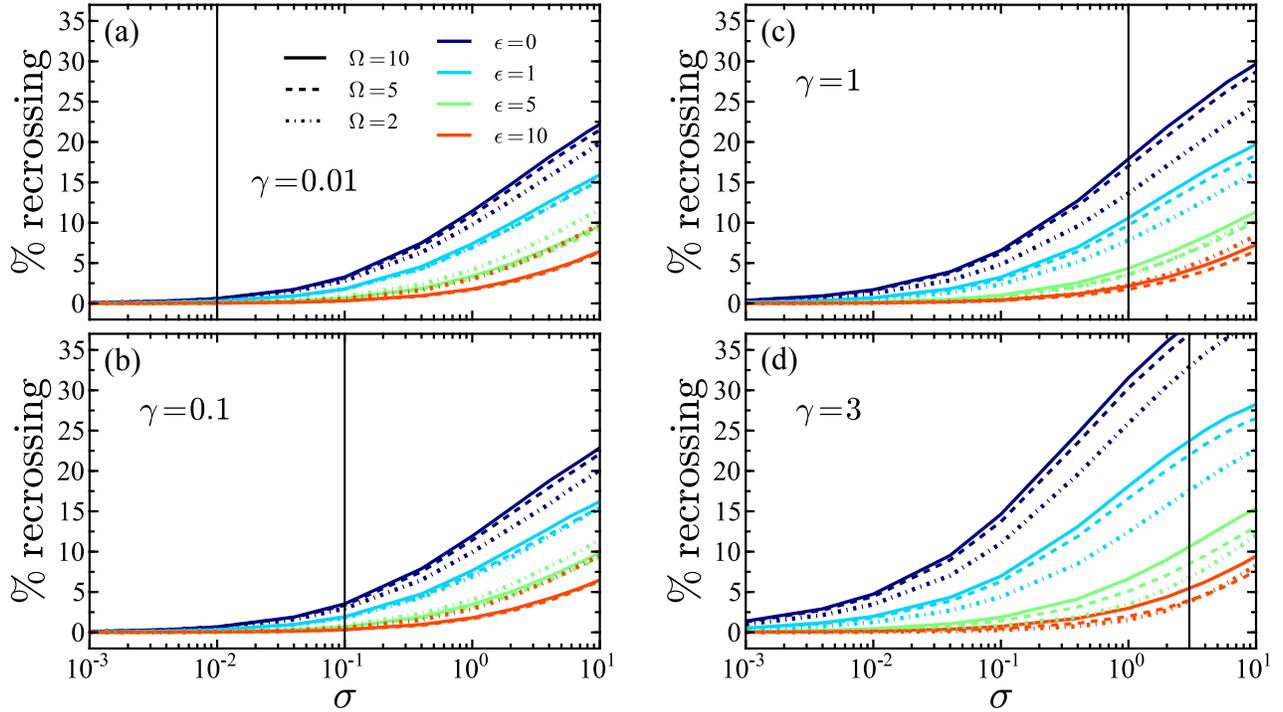}
\caption{\label{fig:recross_1}
The percentage of trajectories that recross the moving dividing surface attached to the DTS~trajectory as a function of
noise strength $\sigma$ with (a) $\gamma=0.01$, (b) $\gamma=0.1$, (c) $\gamma=1$, and (d) $\gamma=3$  
for single-frequency ($\mathbf{\Omega}_1$) driving 
and various values of $\epsilon$ and $\Omega$.
The black vertical line (solid) 
marks the noise strength when the fluctuation-dissipation theorem is obeyed.
}
\end{figure*}
%%%%%%%%%%%%%%%%%%%%%

A comparison between the rates calculated from numerical simulation 
and rates predicted by Eq.~(\ref{eq:rateanhar}) is shown in Fig.~\ref{fig:Flo}.
For all frequency sets $\mathbf{\Omega}_s$ and  parameter values,
agreement is observed.
Underdamped ($\gamma<2$), 
overdamped ($\gamma>2$), and critically damped ($\gamma=2$) regimes of a corresponding harmonic well 
were considered. 
Agreement between the rates persists over all ranges of damping.
For high frequency driving ($\Omega_f > \omb$),
the exponential rate can be averaged over several periods of driving and modulations in the
decay are minimal, as illustrated in Fig.~\ref{fig:Rates}.
Periodic modulations in the decay of $P_\text{R}(t)$ are prominent for low driving frequencies ($\Omega_f \approx \omb$) 
and the integration of $n = 10^8$ trajectories resulted in reaching the numerical asymptote $P_\text{R}(\infty)$ 
at times less than the period of the external driving.
%In cases in which 
In those cases for which the integration time
was insufficient to sample the asymptotic region,
a larger number of trajectories ($n = 10^9$) 
were integrated. 
Each trajectory was integrated to a final time of at least $t_\text{f} = 15$
and, as shown in Fig.~\ref{fig:Rates}, 
$P_\text{R}(\infty)$ is reached 
well before the end of this sampling window.
Increasing the number of trajectories by an order of magnitude
resulted in improved convergence of the scaled logarithmic population and 
marginally better agreement between the compared methodologies,
as shown in Fig.~\ref{fig:Flo}(e) for $\Omega=1$.
The agreement between the two methods is illustrated 
in Fig.~\ref{fig:Flo}(d)
for the smaller, non-unity, driving amplitude case of
$\mathbf{\Omega}_2$.
The decreased driving amplitude leads to a decrease in 
the reaction rate.

%\section{\label{sec:Noise}Deterministic Superimposition on Stochastic Geometries}
\section{Characterizing noisy reactions with the noise-free geometry}
\label{sec:Noise}
%% superimpose is defined as a a combination of objects in which
%% one is placed over another to that both are still evident.
%% I'm not sure that this term readily means that an object that
%% is an undelying representation of the other can be used in its stead
%% As such, I'mn not comfortable using the termp superimposition unless
%% we probide some explanation as to how we are extending its meaning...

%% In the abstract, I suggested earler, "deterministic-limit"
%% How about noise-free geometry?

%%%%%%%%%%%%%%%%%%%%%
\begin{figure}
\includegraphics[width = 8.5cm,clip]{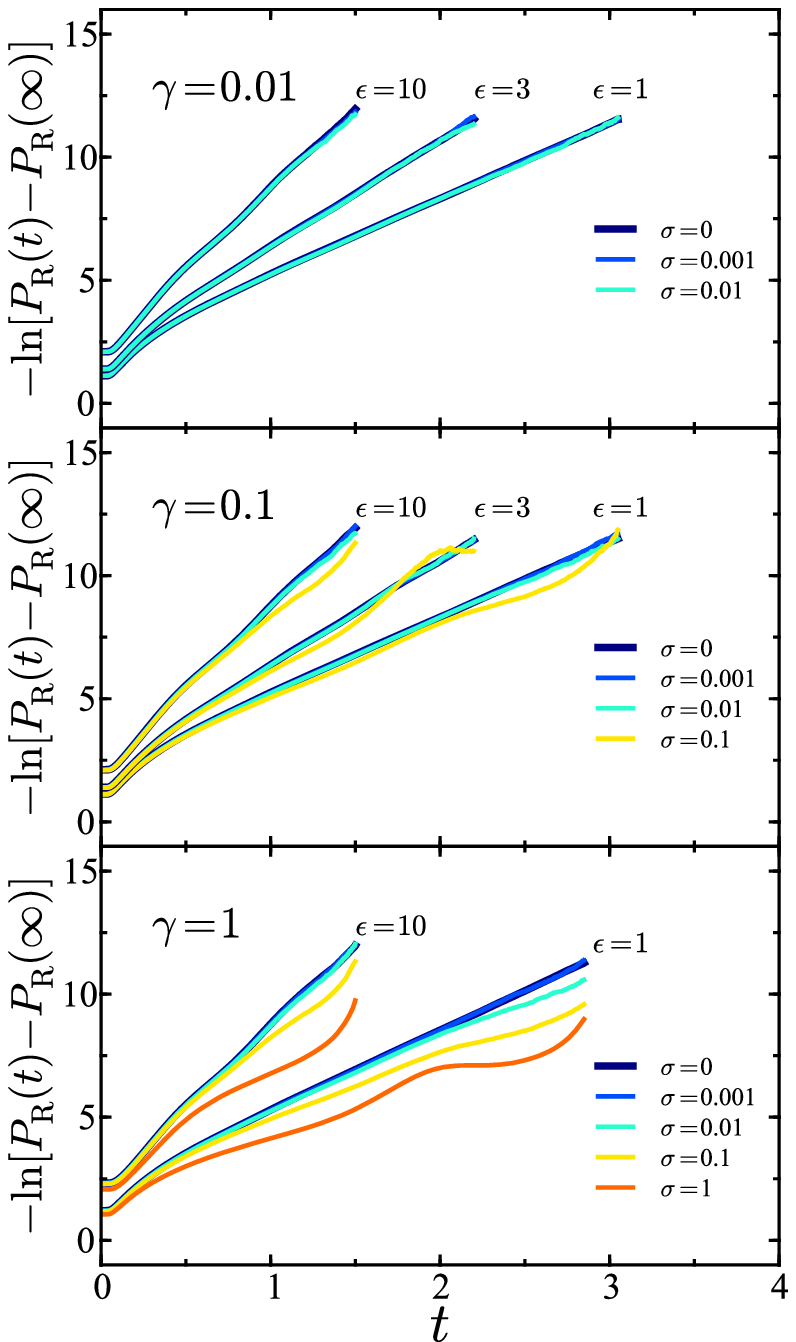}
\caption{\label{fig:Rates_noise}
Time dependence of the scaled logarithm of the reactant population for systems with single-frequency ($\mathbf{\Omega}_1$) periodic and thermal driving
for $\gamma=0.01$ (top), $\gamma=0.1$ (middle), and $\gamma=1$ (bottom). 
The color of each line corresponds to a specific $\sigma$ value.
The decay for systems with various anharmonicites $\epsilon\in{\left\{1,3,10\right\}}$ are shown and denoted in each panel.
The fundamental driving frequency is $\Omega_f=5$ for all panels.
For visual clarity, each curve is truncated at a point 
where the data became noisy.
}
\end{figure}
%%%%%%%%%%%%%%%%%%%%% 

%%%%%%%%%%%%%%%%%%%%%
\begin{figure*}
\includegraphics[width = 17.0cm,clip]{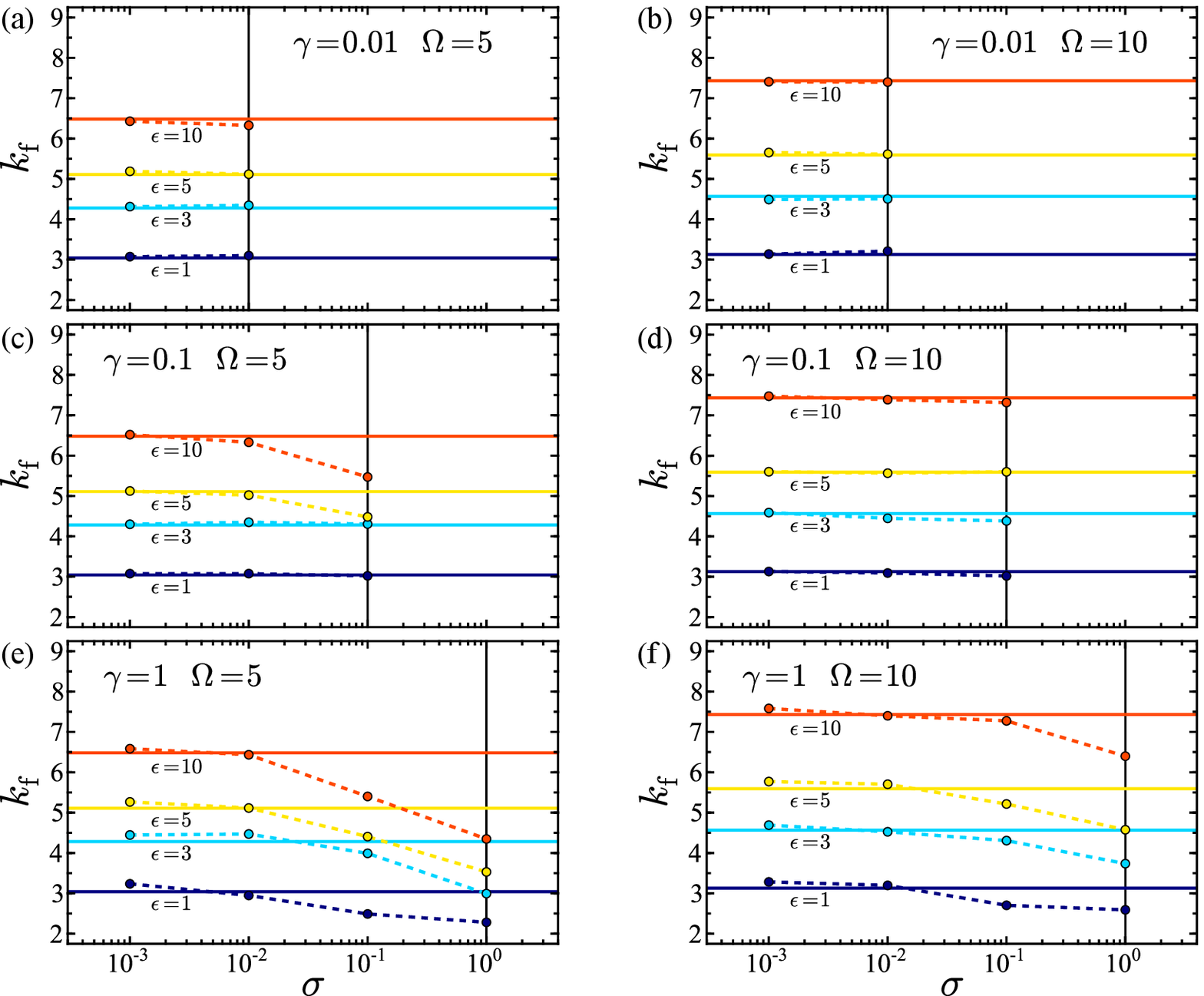}
\caption{\label{fig:thermal_rates}
The barrier crossing rates of systems following the equations of motion~(\ref{eq:motionAnharm}) as a function
noise strength $\sigma$. The rates calculated using the DS attached to the DTS trajectory
for single-frequency ($\mathbf{\Omega}_1$) driving 
and various values of $\epsilon$, $\gamma$, and $\Omega$ are shown as circles.
The horizontal lines (solid) denote the rates given by 
the Floquet exponents of the corresponding DTS trajectory 
and are colored according to a respective $\epsilon$ value.
The black vertical lines (solid) denote the noise strength 
where the fluctuation-dissipation theorem is obeyed.
}
\end{figure*}
%%%%%%%%%%%%%%%%%%%%%

In systems in which the strength of an external driving force 
dominates over that of the thermal driving, 
statistical quantities can be approximated by those of a corresponding purely deterministically driven system.
For thermally induced reactions, Lehmann, Reimann, H{\"a}nggi, and \cite{Lehmann00a,Lehmann00b,Lehmann03} have shown that 
in the overdamped (large-$\gamma$) regime, 
when a chemical reaction is forced by a periodic field the reaction rate is determined in part by the 
geometry of periodic trajectories in the purely deterministic phase space.
This work was later extended to cases with different scaling behaviors between the 
strength of thermal activation and the strength of the external field.\cite{Maier01,Dyk04,Dyk05}

Our goal here is to develop a minimalist theory,  
applicable at the limit where the magnitude $\sqrt{2 \sigma}$ 
of a noise sequence $\xi_\alpha(t)$ 
is a small enough perturbation to the periodic driving $E(t)$ that 
%by superimposing the TS trajectory on to the noisy geometry a DS with minimal recrossings can be obtained.
the TS~trajectory of the noiseless system (the periodic orbit) gives rise to a DS with minimal recrossings.
This deterministic TS~trajectory (DTS~trajectory) does not solve the equations of motion~\eqref{eq:motionAnharm} with a non-zero value of $\sigma$.
We therefore distinguish the DTS~trajectory from the true TS~trajectory of the 
noisy system
(that we do not compute in this work).
%Note that due the inclusion of the term $\sqrt{2 \sigma}\xi(t)$ 
%with non-zero value of $\sigma$ in Eq.~(\ref{eq:motionAnharm}),
%the TS trajectory is a fictitious object and is not a solution to the equations of motion.
%We therefore will differentiate the true TS~trajectory from the superimposed deterministic transition state (DTS).

%A principal assumption for application of deterministic superimposition 
A principal assumption for the use of the noise-free geometry
is that the phase space density of the thermal system, and its time-dependence, 
is approximately that of the deterministic system, i.e., 
$p_t(\Delta x_\alpha,\Delta v_\alpha) \approx p_t(\Delta x,\Delta v)$. 
As shown in Fig.~\ref{fig:Phase}, 
for small values of $\sigma$ the geometry of the thermal system is  similar to that of its deterministic counterpart.
%For periodically driven phase space density evolution, 
%the DTS trajectory with corresponding Floquet exponents can be readily obtained and thus the rate theory 
The rate theory developed in Sec.~\ref{sec:Floquet} 
for the deterministic system can therefore be applied.
This is advantageous in applications such as in
comparisons with experiments in which the exact noise sequence is not known.

Thermal systems in which the fluctuation-dissipation theorem (FDR) 
is not obeyed due to energy dissipation constitute non-equilibrium processes.
Formal treatments of fluctuation-response in periodically forced systems by Teramoto, Harada, and Sasa \cite{Sasa05,Teramoto05} 
provide insight into the rate of energy dissipation in such systems. 
Green \textit{et al}.\cite{Green2013} have shown that the rate of energy dissipation is directly related to the dynamical entropy of the system. 
To realize non-equilibrium conditions in the present model reaction, 
the damping constant $\gamma$ is held constant
% the shrinkage rate of the phase space volume is held constant $e^{-\gamma t}$ 
and the strength of the thermal fluctuations $\sigma$ is increased up to 
the point where the FDR is satisfied.
If the initial velocities are drawn from a Boltzmann ensemble 
with $\kT=1$ (in dimensionless units), this is the case at $\sigma=\gamma$.
If $\sigma<\gamma$ the thermal bath is at a lower temperature than that of the distribution of initial velocities.
In this case, the temperature of the ensemble of reactants will be continually cooled by the thermal bath.
Thermal annealment of the ensemble calls for the development of postmodern rate theories 
which rely singularly on geometric properties of phase space, and not the dynamics of the ensemble itself. 
This is in stark contrast to the TST assumption of
equilibrium distributions in metastable states and at the TS.

The percentage of thermal trajectories that recross the DS attached to the DTS trajectory 
is shown in Fig.~\ref{fig:recross_1} for varying noise strengths $\sigma$
and constant dissipation rates.
As shown in  Figs.~\ref{fig:recross_1}(a) and \ref{fig:recross_1}(b),
a minimal number of recrossings occur below and up to the 
FDR threshold for small values of $\gamma$.
For the $\gamma=1$ case, shown in Fig.~\ref{fig:recross_1}(c), 
trajectories persist around the BT for long times, leading to a larger number of recrossings than observed for
smaller dissipation rates.
For the overdamped dynamics ($\gamma=3$), shown in Fig.~\ref{fig:recross_1}(d), 
the deterministic~DS identifies reactive trajectories adequately
only for weak thermal driving (small $\sigma$) and strong anharmonicity. 
As the harmonic limit is approached or in equilibrium systems the superimposed DS becomes very poor.

The decay of the scaled logarithm of the normalized reactant population, 
as calculated with the superimposed deterministic DS,
is shown in Fig.~\ref{fig:Rates_noise} for various parameter values.
Over all friction regimes,
the population decay of the systems with additional thermal driving follows 
that of its deterministic counterpart 
if the noise strength $\sigma$ is sufficiently low.
For $\gamma=1$, when the strength of the thermal driving approaches that of the FDR, a decrease in the reaction rate is observed.
The data presented in Fig.~\ref{fig:Rates_noise} becomes highly oscillatory at long times due to recrossings of the DS. 
For visual clarity each data series has been truncated to remove this noisy tail. 
As observed in Figs.~\ref{fig:Rates} and \ref{fig:Rates_noise}, for short times ($t \lessapprox 0.3$), the decay of $P_\text{R}$
is non-exponential, signifying temporally global non-RRKM kinetic behavior. 
We obtain the rate from the long-time asymptotic decay of 
of $P_\text{R}$, 
which is representative of kinetic experiments in which the concentration 
of a reactant species is directly measured over time.\cite{Levinepchem}

The thermal rates calculated using the DTS trajectory are shown in Fig.~\ref{fig:thermal_rates}. 
As expected by the minimal number of recrossings shown in 
Fig.~\ref{fig:recross_1}, stability analysis of the DTS can produce an excellent approximation to the rate in thermal environments. 
Through calculation of the error between the numerically calculated rate with included noise and the 
rate given by the Floquet exponents of the DTS trajectory, 
the extent of applicability of the noise-free geometry can be quantified.
%the extent of applicability of superimposition can be quantified.
%%% superimposition!? -- HERE
This error is $< 3\%$ at $\gamma=0.01$ over all parameter values.
It is $ < 1\%$ for $\epsilon \in \left\{5,10\right\}$.
Increasing the dissipation by an order of magnitude ($\gamma=0.1$) results 
in the same general trends, with all errors 
generally less than $5\%$.
The exceptions occur at the noise strength 
where the FDR is obeyed ($\sigma=0.1$) 
at $\epsilon \in \left\{1,3\right\}$ and $\Omega=5$ 
for which the error $\approx 20\%$. 
For $\gamma=1.0$ and $\Omega = 10$, 
all calculated errors are less than or on the order of $20\%$, 
increasing monotonically as a function of $\sigma$.
As illustrated in Fig.~\ref{fig:thermal_rates}(e), 
at lower-frequency driving ($\Omega = 5$) and large noise ($\sigma=1$), 
the error is between $30\%-50\%$.
This suggests a practical upper bound to the applicability of 
the noise-free geometry in estimating the reaction rates in the
presence of noise.
%deterministic superimposition to estimate reaction rates.
Although not shown, for overdamped dynamics,
stability analysis of 
the DTS gives an accurate approximation to the rate
only in non-equilibrium small noise regimes.

The calculated errors are on the order of the error expected from 
application of variational transition state theory (VTST).\cite{truh84} 
The presented methodology is advantageous over VTST as it does not require the integration of large numbers of trajectories or a flux minimization procedure.
Thus, stability analysis of the DTS trajectory offers a simple rate calculation methodology that can be readily applied, in weak thermal environments,
to driven chemical reactions with only prerequisite knowledge of the geometry of the energy surface and the
functional shape of the driving waveform.

\section{Conclusions}\label{sec:Conclusions}

We have shown that in a model chemical reaction
subjected to the influence of forcing from a temporally
periodic external field, a recrossing-free dividing surface
can be constructed over an unstable periodic orbit in the
region of a moving energetic barrier top. This forms the
basis for future work on specific driven chemical reactions
that can be represented by a single collective variable for
the reaction coordinate under nonequilibrium conditions.
Potential targets include both substitution and isomerization
reactions in which the governing multi-dimensional energy
surface can be parameterized by a single collective degree
of freedom. Other possible targets include mechanochemical
reactions and stimuli-responsive assembly mechanisms when
the reaction rate is dictated predominantly by geometric
properties about the moving dividing surface. Generally,
force-modified and temperature-modulated energy surfaces,
deforming under the influence of temporally varying forces,
motivate the development of rate methodologies that go
beyond the simplistic equilibrium arguments intrinsic to
classical equilibrium transition state theory.

The no-recrossing surface constructed here has been
shown to persist for strongly anharmonic barriers subjected to
single-mode and multi-mode driving waveforms. A formally
exact rate theory has been developed based on the flux of
reactive trajectories through this recrossing-free surface. It
rectifies the principal criterion of transition state theory for
periodically driven chemical reactions.

To circumvent computationally taxing numerical calculations
of the reactive flux through this surface, a rate theory has
been developed based on the stability of the dividing surface.
Strong agreement was observed between the rate predicted
by the Floquet exponents of a trajectory defining the phase
space evolution on the dividing surface, and the rate calculated from
simulation of a large ensemble of trajectories. Thus, in a
periodically driven chemical reaction, the asymptotic decay
rate of an initial distribution of reactants can be extracted
directly from the stability of the time-varying dividing surface
irrespective of the dynamics of the reactive population.

Use of the noise-free geometry to approximate the
corresponding structure of a driven thermal system has been
shown to give an excellent approximation to the optimal
dividing surface if the magnitude of the oscillating force is
large compared with that from the thermal environment. For
thermally activated processes, the stability exponents of the
purely periodically driven system can thus be used to predict
the reaction rates without an explicit treatment of the thermal
dynamics. Extensions of the this work to include an explicit treatment of the noise, including
systems with structured solvents environments \cite{hern13b, hern14d} and systems 
displaying fluctuating rates \cite{Green2014} are possible next steps,
and ones which we are currently pursuing.

\section{Acknowledgments}\label{sec:thanks}
This work has been partially supported by the Air Force 
Office of Scientific Research through Grant No.~FA9550-12-1-0483.
%This work has been partially supported by the National Science Foundation (NSF)
%through Grant No.~NSF- CHE-1112067.
Travel between partners was partially supported through the People Programme (Marie Curie Actions)
of the European Union's Seventh Framework Programme FP7/2007-2013/ under REA Grant Agreement No. 294974.

%%%%%%%%%%%%%%%%%%%%%%
%\bibliography{DrivenAndNoise}
\bibliography{j,hern,gas,tst,osc-bar,halcyon}
\newpage
\printtables
\newpage

\end{document}